%% file: main.tex
\documentclass[sigconf,nonacm, 9pt]{acmart}
\usepackage{subfigure}
\usepackage{multirow}
\usepackage{makecell}
\usepackage{algorithm}
\usepackage[algo2e,linesnumbered,ruled]{algorithm2e}
\AtBeginDocument{%
  \providecommand\BibTeX{{%
    \normalfont B\kern-0.5em{\scshape i\kern-0.25em b}\kern-0.8em\TeX}}}

\newcommand{\para}[1]{\smallskip\noindent {\bf #1}}

\long\def\Wei#{}  
\long\def\Lili#{}
\long\def\Note#{}

\long\def\comment#1{}
\newcommand{\ie}{{\em i.e.}}
\newcommand{\eg}{{\em e.g.}}
\newcommand{\ours}{{\em MFB }}

\begin{document}
\author{Wei Sun, Lili Qiu}
\affiliation{
\institution{The University of Texas at Austin}
\country{USA}
}
\email{{weisun, lili}@cs.utexas.edu}
\title{Visual-Assisted Sound Source Depth Estimation in the Wild}

\input{abs}


\maketitle

\input{intro}
\input{bg}

\input{bg2}

\input{approach2}

\input{eval}

\input{related}
\input{limit}

\input{conclusion}


\bibliographystyle{ACM-Reference-Format}
\bibliography{tracking,sample-base}

\end{document}

%% file: abs.tex
\begin{abstract}

Depth estimation enables a wide variety of 3D applications, such as robotics, autonomous driving, and virtual reality. Despite significant work in this area, it remains open how to enable accurate, low-cost, high-resolution, and large-range depth estimation. 
Inspired by the flash-to-bang phenomenon (\ie hearing the thunder after seeing the lightning), this paper develops FBDepth, the first audio-visual depth estimation framework. It takes the difference between the time-of-flight (ToF) of the light and the sound to infer the sound source depth. FBDepth is the first to incorporate video and audio with both semantic features and spatial hints for range estimation. It first aligns correspondence between the video track and audio track to locate the target object and target sound in a coarse granularity. Based on the observation of moving objects' trajectories, FBDepth proposes to estimate the intersection of optical flow before and after the sound production to locate video events in time. FBDepth feeds the estimated timestamp of the video event and the audio clip for the final depth estimation. We use a mobile phone to collect  3000+ video clips with 20 different objects at up to $50m$. FBDepth decreases the Absolute Relative error (AbsRel) by 55\% compared to RGB-based methods. 
\end{abstract}

%% file: intro.tex
\section{Introduction}
\label{sec:intro}

Depth estimation has been a popular topic owing to many important applications, including autonomous driving, robotics, medical diagnosis, 3D modeling, augmented reality, and virtual reality (VR). With the depth information, 2D RGB images can be mapped into 3D space to enable a wide variety of 3D applications as well as boost many computer vision tasks. 

Owing its importance, there have been significant efforts on developing depth estimation techniques using LiDAR~\cite{nuscenes, oxford},  structured-light~\cite{kinect}, RGB cameras, ultrasound (\eg, ~\cite{CAT,llap,Strata,wang2019millisonic}), WiFi (\eg, ~\cite{vasisht2016decimeter,WiDeo}), and mmWave (\eg, ~\cite{mTrack}). Despite significant work, how to achieve high accuracy, large range, low cost, and wide availability remains an open problem.

In this paper, we develop a novel passive depth estimation scheme, called Flash-to-Bang Depth (FBDepth). 
It is inspired by a well-known phenomenon -- "flash-to-bang", which is used to estimate the distance to the lightning strike according to the  difference between the arrival time of a lightning flash and a thunder crack. This works because light travels a million times faster than sound, When the sound source is at least several miles away, the delay difference is large enough to be measured.

Applying it to our context, we can estimate the depth of a collision that triggers both audio and visual events and use the arrival time difference between the video event and audio event to calculate the distance. Collisions take place in many contexts when a ball bounces on the ground, a person steps on the floor, and a musician plays an instrument (\eg, drum, piano). Therefore, FBDepth can be applied to many scenarios, such as sports analytics and surveillance. FBDepth can be combined with other depth estimation schemes, including monocular depth estimation, to further enhance the accuracy. 



We decompose the depth estimation problem into the following three important steps: (i) identifying the object in the video that generates the collision sound in the audio recording, (ii) detecting the collision time in the video recording, and (iii) detecting the collision time in the audio recording and using the time difference to estimate the depth.  

\begin{figure}[h]
\centering
\includegraphics[width=0.48\columnwidth]{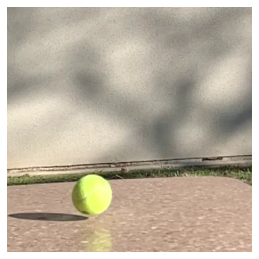}
\includegraphics[width=0.48\columnwidth]{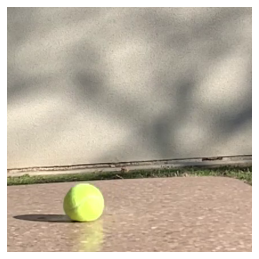}
\caption{An example collision in a video recording.\Wei{put these two figures into the background and relationship between the audio and video event section}}
\label{fig:video-demo}
\end{figure}

\begin{figure}[h]
\centering
\includegraphics[width=0.98\columnwidth]{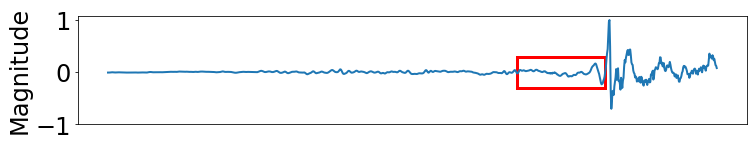}
\includegraphics[width=0.98\columnwidth]{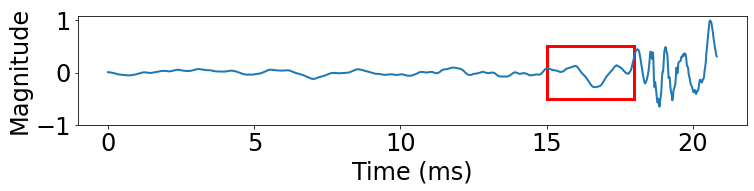}
\caption{An example collision in an audio recording.}
\label{fig:audio-demo}
\end{figure}

While these steps appear intuitive, it is challenging to realize an accurate depth estimation system on a commodity mobile device for the following reasons. First, we need to automatically associate the object that generates sound in the audio with that in the video, and generalize to diverse scenarios. Second, as illustrated in Figure~\ref{fig:video-demo}, even at the highest frame rate available on smartphones (\eg, 240 frames per second (FPS)), the collision event may not be captured in the video (\eg, the ball has not touched the surface in one frame but deformed in the next frame). Moreover, even with super high frame rates beyond 240 FPS, depending on the viewing angle, it could be challenging to tell exactly when the ball just touches the floor. Third, detecting the event in the audio recording is also challenging. Figure~\ref{fig:audio-demo}(a) shows that the collision generates a strong impulse immediately upon touching the surface while Figure~\ref{fig:audio-demo}(b) shows that the collision generates a soft sound at first and then increases to a loud sound. It is challenging to automatically label the collision time in the audio recording manually for supervised learning. Note that 1 ms detection error translates to 34 cm distance error. Therefore, in order to achieve within 50 cm error, the audio and video event detection error should be capped within 1-2 ms, which is challenging because audio and video both have limited sampling rate and audio and video may not start recording at the same time.

We approach these challenges as follow. First, we take the audio recording and a video frame as the inputs to identify the pixels belonging to the object that produces the sound. We leverage SoundNet~\cite{SoundNet} to extract audio features and ResNet~\cite{ResNet} to extract visual features, and then train a new neural network to derive the audio video alignment (\ie, which pixels in the video frame correspond to the object that generates the collision sound). This output serves as an attention map in the following steps so that we can focus on the target object and remove irrelevant regions from the video frame. 
To further enhance its performance, we are only interested in the object whose movement causes a collision, so we use the optical flow to filter out irrelevant static pixels. This not only speeds up computation but also improves the accuracy. 

\Wei{emphisize using optial flow can be more fine grained to detect the intersection}
Next, we detect the collision event in the video. Since it is challenging to manually label the collision time in the video, we develop a way to automatically derive the collision time. Our approach consists of two phases: (i) coarse-grained detection and (ii) fine-grained detection. In (i), we leverage our observation that an object changes its moving direction upon collision. Therefore, we can use optical flow to approximate the motion field and detect the collision based on a significant change in the optical flow. This yields a rough estimate of the collision time. To get a more precise collision time, in (ii) we use a few frames before and after the collision time to estimate the collision time by finding the intersection between the movement trajectories before and after collision. Note that during this step, instead of interpolating the video frames, we interpolate the motion trajectory, which allows us to achieve high accuracy and efficiency. 


Finally, we analyze the audio recording to identify the collision event in the audio and derive the distance estimate. Note that we combine audio event detection with the distance estimate using one neural network to remove the need of manually labeling the collision time in the audio and directly train using the ground truth distance, which is easy to obtain. We detect the collision event in the audio by designing a deep neural network to analyze the audio samples around the collision time where the audio samples very close to the video collision time are highlighted. We train a neural network that outputs the estimated depth directly based on the difference between the audio and video event time. 

Our contributions can be summarized as follow:

\Wei{
1. To the best of our knowledge, FBDepth is the first passive audio-visual(AV) depth estimation. It uses widely available sensors -- a camera and a microphone to estimate the range based on the flash-to-bang phenomenon. 

2. We develop a systematic audio-visual framework by modeling AV semantic correlation and spatial correlation simultaneously. It includes a motion-aware AV alignment to localize sound sources in the video and audio, a novel coarse-grained to fine-grained video event detection by leveraging the motion change and consistency, and a depth estimation network for audio samples augmented by video estimation. 

3. We collect 3k audio-visual samples across 20 different objects. Our extensive evaluation shows that FBDepth achieves 0.69m absolute error(AbsErr) and 4.7\% AbsRel across a wide range of distances (\eg, 2 m -- 50 m).
}

The rest of this paper is organized as follows. 
We present our approach in Section~\ref{sec:approach}. We describe evaluation results in Section~\ref{sec:eval}. We survey related work in Section~\ref{sec:related}. Finally we conclude in Section~\ref{sec:conclusion}. 


\comment{
%
While intuitive, it is challenging to realize an accurate depth estimation system on a commodity mobile device due to the mismatch between the millisecond temporal resolution and the incompetence of audio visual data collection and interpretation in the wild. This challenge applies to many contexts in the system. One key problem is caused by multiple objects and sounds coexisting in the same temporal range. It confuses the system to associate the corresponding object and its special sound and leads to the mismatch of video event timestamp and audio event timestamp. Thus, we design the motion based audio visual alignment model to generate the attention to the target object as well as extract the audio clip of best fit. We feed different pairs of audio clips and frames together with optical flows to the model. The positive pair is consisted by a frame including the moving object and a audio clip having the start of object sound. The other pairs are negative. The model can predict an attention map to focus on the target object and a binary discrimination on whether the pair is positive. It helps to remove unrelated visual cues and localize audio event at a coarse granularity.

Figuring out the video event timestamp is extremely difficult in intuition. In fact, it is unable to labelling the event timestamp in the video with sub-millisecond resolution with human effort. The limited video frame rate ranging between 30 -- 240 frames per second (FPS) cannot capture the collision beginning moment. Figure~\ref{fig:video-demo} shows two successive frames in 240 FPS recording. The ball has not touched the surface in the first frame but deformed in the second frame. Besides, it is not easy to apply super high frame rate to distinguish the exact frame at which the ball touches the surface by only observing the visual cues because of the blurry pixels and the ill-suited angle of the view.  Based on the misfortune of no ground truth timestamp, it becomes more difficult to build an effective estimation model. To resolve this dilemma, we propose a novel motion intersection based on two individual consistent motions before and after the collision. We observe that the collision moment is the final status of the motion before the collision as well as the initial status of the motion after the collision. We adopt the optical flow to capture the motion consistency as pixel level and predict the position and pose of the object between successive frames. Finally, the most similar moment is estimated as the intersection of the motion before and after the collision. This scheme can predict and interpolate the unseen collision moment from historical frames.

The labelling problem exists in the audio track as well even though the sampling rate of audio is sufficient. Figure~\ref{fig:audio-demo} shows the impact sound from different objects in the wild. Figure~\ref{fig:audio-demo}(a) shows that the collision generates a strong impulse immediately upon touching the surface while Figure~\ref{fig:audio-demo}(b) shows that the collision generates a soft sound at first and then increases to a loud sound. The key observation is that various sound production makes it impossible to map the audio signal features exactly to the corresponding collision moment with human effort. We cannot sense the sound when a ball is just touching the reflection surface. Thus, there is no reliable ground truth  in the audio track to differentiate from the video event timestamp. Apart from this physical audio visual uncertainty, there is a non-trivial unsynchronization due to variable delay between camera recording and microphone recording. Consider all these variable offsets, we propose to avoid estimating audio event timestamp and estimate the depth directly from the featured audio track. We insert the video event timestamp in to the audio track and enable the model to adapt the offsets based on different signal pattern. Instead of the Short-time Fourier transform (STFT), the 1D convolutional encoder is adopted to capture fine grained features in the temporal domain.

Our contributions can be summarized as follow:

\begin{itemize}
    \item To the best of our knowledge, this is the first work that leverages the "flash-to-bang" phenomenon for depth estimation. It complements existing depth estimation as it does not need special hardware or calibration to a new environment. 
    \item We design an innovative audio-visual learning framework to leverage audio visual cues from semantic context and physical characteristics. It shows the capability to handle the background noise and interference as well as uncertain offsets. 
    \item We develop a novel video event detection with millisecond temporal resolution by leveraging the intersection of motion trajectories before and after collision.
    \item Our extensive evaluation shows that XXX.
\end{itemize}

The rest of this paper is organized as follow. It overviews the system architecture in Section~\ref{sec:overview}, followed by expanding on these contributions in Section~\ref{sec:approach}, and evaluation results in Section~\ref{sec:eval}.

}

%% file: bg.tex
\section{Background on Depth Estimation}

\Wei{In this section, we survey a variety of depth estimation approaches and put FBDepth into perspective by comparing with the other approaches using several different metrics.}



As shown in Table~\ref{tab:category}, we classify the existing approaches broadly into two main categories: active sensing and passive sensing. Active sensing methods actively emit signals, such as LiDAR, structured-light, radar, ultrasound, WiFi, RFID, or mmWave. They use time-of-flight (ToF), amplitude, phase or Doppler shift to estimate the range. Passive sensing uses signals from the environment for sensing. It commonly uses a monocular camera~\cite{monocular1} or a stereo camera~\cite{zed2} by leveraging the implicit monocular or binocular depth clues from images. 

\para{Accuracy:} 
Active sensing can achieve centimeter-level accuracy. LiDAR is one of the most popular methods because it can generate dense point clouds and achieve good accuracy. 

Many passive sensing schemes use deep learning for range estimation. Monocular camera based depth estimation can achieve within 10\% AbsRel up to 80 m~\cite{monocular2}. However, its absolute error (AbsErr) increases linearly with the distance. The AbsErr can be several meters when objects are tens of  meters away. Moreover, it relies on high-quality datasets and cannot handle arbitrary scenes and fields of view, which may not contain enough clues for depth estimation.  
The performance of the stereo camera based depth estimation depends on the separation between two cameras~\cite{HyperSight} (\eg, its AbsRel is ~5\% on camera Zed 2). Its AbsErr increases faster than a linear rate in the RGB camera based method.



\begin{table*}[h!]
\setlength{\tabcolsep}{2pt}
\centering
\vspace*{-0.1in}
\begin{tabular}{ c| c | c | c | c | c | c | c}
    \toprule
    Mode & Sensor & Device/Method & Accuracy & Range & Angular Resolution & Power & Extra Cost \\
    \hline
    \multirow{7}{5em} {Active}
    & LiDAR            & Velodyne HDL-32E~\cite{oxford}        & 2cm & 100m & \makecell{vertical: 1.33°\\ horizontal: 0.1° to 0.4°} & 10W & >\$5K\\
    & structured light & Realsense D455   & 2\% & 6m & pixel-level & USB charge & \$ 400 \\
    & ToF camera    & Azure Kinect~\cite{kinect} & < 1cm & 6m & pixel-level & USB charge & \$ 600 \\
    & mmWave      & Navtech CTS350-X~\cite{oxford}   & 4.38cm & 163m & 1.8° beamwidth & 20w & > \$ 500 \\
    & inaudible sound             & Rtrack~\cite{Rtrack}   & ~2cm & 5m & > 10° & on-device & 0 \\
   & WiFi  & Chronos~\cite{vasisht2016decimeter}& 65--98 cm & $<50 m$&  & Intel 5300 & $<\$50$ \\
    \hline
    \hline
    \multirow{5}{5em}{Passive}
    & camera               &      MonoDepth2 ~\cite{monocular2}   & 10.6\% & 80m(Kitti dataset) & pixel-level & on-device & 0\\ 
    
    & stereo camera &Zed 2 camera~\cite{zed2}    & \makecell{< 1\% up to 3m; \\ < 5\% up to 15m} & 20m & pixel-level & USB charge & \$ 450 \\
    & camera + mic  & FBDepth    & \makecell{overall 4.7\%; \\ > 30m: 3.1\% } & 50m & depth for each obj. & on-device & 0 \\
    \hline
\end{tabular}
\caption{A comprehensive comparison for different depth estimation approaches. Among these metrics, it is better for accuracy, range and resolution to be high while it is promising for power and cost to be low.}
\label{tab:category}
\vspace*{-0.2in}
\end{table*}

\para{Range:} LiDAR and radar for autonomous driving can achieve up to hundreds of meters. Other indoor estimation methods  can achieve an effective range of about 5m.
Monocular camera based depth estimation can achieve 1000 meters~\cite{farsight}. 
Stereo camera based estimation is constrained by the separation between the two cameras. Zed 2 can support up to 15m. 
FBDepth can support 50 m. Its large range comes from two major factors: (i) it uses directly received audio signals instead of reflected signals and (ii) it uses audible frequencies, which has a slower decay and stronger frequency response than inaudible audio frequencies. 

\para{Resolution: } The resolution refers to the granularity that each depth measurement point corresponds to the 3D space. RGB-based methods, ToF camera, and structured light can achieve pixel-level estimation. RF-based and acoustic based solutions can only detect sparse reflection points. The LiDAR with multiple beams can generate a dense point cloud,  the density of its point cloud decreases quadratically with the distance~\cite{density} and its point cloud becomes pretty sparse at a large distance. As a result, autonomous driving datasets commonly annotate point clouds up to 80m~\cite{nuscenes, kitti}. 

\para{Power: }
Active sensing methods need to emit modulated signals, which tend to consume more power in order to get a reasonable SNR from the reflected signal over a large detection range. When they are applied to mobile phones, their performance is constrained by the power. For example, iPhone Pro has a Lidar on the back of the phone with a max range of 5m and a ToF camera on the front with a max range of 50cm. 

\para{Cost: } Cameras and microphones are widely available. In comparison, LiDAR is much more expensive and not available on most mobile devices. mmWave also has limited deployment. While WiFi is popular, in order to achieve high accuracy, we need more spectrum and PHY layer information, both of which limits its availability on mobile devices. 
FBDepth only requires a camera and a microphone, which makes it possible for wide deployment. 


\para{Summary:} There are many depth estimation technologies. Our FBDepth complements the existing solutions by adding a low-cost, easy-to-deploy, accurate, and long range solution. It has higher accuracy than existing monocular camera-based solutions and a longer range than existing stereo camera-based solutions.


\comment{
\section{Audio-visual Modeling}

\Wei{In this section, we describe the characteristics of audio and visual events and how we can correlate them in our context.}

\subsection{Problem Formulation}

We start with the fundamental principle of the audio-visual depth estimation. The flash-to-bang exploits the significant difference in the propagation delays of the light and sound. Specifically, we have 

\begin{equation}
\frac{d}{v} - \frac{d}{c} = T
\label{eq:flash-to-bang}
\end{equation}
where the depth of the AV event is $d$ and the difference between the TOF of sound and light is $T$. $c$ and $v$ denote the propagation speeds of light and sound, respectively. We can estimate $d$ based on $d = \frac{cvT}{c-v} \approx vT $ since $c \gg v$. 

To measure $T$, we observe $T=T_{audio} - T_{video} + T_{hardware}$, where $T_{audio}$ and $T_{video}$ denote the event time in the audio and video recordings, respectively, and $T_{hardware}$ denotes the start time difference in the audio and video recordings. We implicitly learn and compensate for $T_{hardware}$ when we try to match our estimated distance with the ground truth distance. 
\Note{note:hardware delay can be tuned in the low level API to keep a small variance and be learned implicitly; explicitly estimate the T video}
\Note{note: 240 fps and 48khz is common on many devices; some can achieve even 960 fps and 96khz. temporal resolution is good from raw data}

Human can only sense the delay more than $100 ms$~\cite{voip} to derive a rough depth estimate for the event (\eg, with an error in tens of meters). Thanks to advance in sensors on mobile devices (\eg, up to 240 FPS cameras and at least 48 KHz microphones), we have the potential to achieve decimeter-level depth estimates.



\subsection{Audio-visual Semantic Correlation }
\Wei{Audio-visual learning has become a popular trend in computer vision, speech, and robotics. Recent work takes advantage of the coexistence of audio events and video events to enhance the performance of analytic tasks and enable novel applications. They focus on the semantic correlation between the two modalities and associate visual representations and audio representations. 

In this paper, we are interested in a specific set of audio-visual events where the impact sound is triggered by a collision. These events are common in nature. For example, an object is freely falling or bounced by a racket or wall, or a person is stepping on the floor. They also follow semantic correlation. Each object produces a unique sound during the collision and has a unique semantic correlation. Only the moving objects with an impulse force can generate the impact sound. The two folds of semantic correlation bring opportunities to simultaneously analyze the audio and video data.  }

\subsection{Audio-visual Spatial Correlation}
We introduce a novel audio-visual spatial correlation in this work. We sense the delay between the audio event and the video event based on the difference between the propagation speeds of light and sound. The delay is perceptible if the event happens far away, such as a lightning strike, a flying plane, and a boom. However, it is hard for humans to estimate the range if the audio-visual event is close by.  

Thanks to the advance in sensors, many mobile devices can support cameras with 240 FPS and even up to 960 FPS and microphones with 48 kHz sampling rate and up to 96 kHz). The high audio/video sampling rates make it possible to accurately measure the delay. The sensing scheme can leverage audio-visual events to enable passive ranging. 

We observe that the collision lasts at most 16ms (\eg, 4 frames in 240 FPS). The impact sound is sometimes even shorter (\eg, 10 ms). 
Such a short collision makes it highly unlikely to have multiple collisions occur exactly at the same time. Therefore, it is reasonable to assume the impact of sound from different collisions does not overlap in time. If these impact sounds do overlap, there are well-known sound source separation algorithms that can be used to separate these impact sounds. On the other hand, our system allows multiple objects that generate collisions to appear in the same video by first identifying all objects that cause collisions in the video and then applying the same method to each object for range estimation.



}

%% file: bg2.tex
\section{Overview}

In this section, we describe the characteristics of audio and visual events and how we can correlate them in our context.

\subsection{Problem Formulation}

We start with the fundamental principle of the audio-visual depth estimation. The flash-to-bang exploits the significant difference between the propagation delays of the light and sound. Specifically, we have:

\begin{equation}
\frac{d}{v} - \frac{d}{c} = T
\label{eq:flash-to-bang}
\end{equation}
where the depth of the AV event is $d$ and the difference between the TOF of sound and light is $T$. $c$ and $v$ denote the propagation speeds of light and sound, respectively. We can estimate $d$ based on $d = \frac{cvT}{c-v} \approx vT $ since $c \gg v$. 

To measure $T$, we observe $T=T_{audio} - T_{video} + T_{hardware}$, where $T_{audio}$ and $T_{video}$ denote the event time in the audio and video recordings, respectively, and $T_{hardware}$ denotes the start time difference in the audio and video recordings. We implicitly learn and compensate for $T_{hardware}$ when we try to match our estimated distance with the ground truth distance. 

Humans can only sense the delay of more than $100 ms$~\cite{voip} to derive a rough depth estimate for the event (\eg, with an error in tens of meters). Thanks to advances in sensors on mobile devices (\eg, up to 240 FPS cameras and at least 48 KHz microphones), we have the potential to achieve decimeter-level depth estimates.

\subsection{Audio-visual Semantic Correlation }
\Wei{Audio-visual learning has become a popular trend in computer vision, speech, and robotics. Recent work takes advantage of the coexistence of audio events and video events to enhance the performance of analytic tasks and enable novel applications. They focus on the semantic correlation between the two modalities and associate visual representations and audio representations. 

In this paper, we are interested in a specific set of audio-visual events where the impact sound is triggered by a collision. These events are common in nature. For example, an object is freely falling or bounced by a racket or wall, or a person is stepping on the floor. They also follow semantic correlation. Each object produces unique sound during the collision and has unique semantic correlation. Only the moving objects with an impulse force can generate the impact sound. Therefore, we leverage the motion from the video along with the impact sound from the audio for range estimation.}

\subsection{Audio-visual Spatial Correlation}
\label{ssec:av-spatial}
We introduce a novel audio-visual spatial correlation in this work. We sense the delay between the audio event and the video event based on the difference between the propagation speeds of light and sound. The delay is perceptible if the event happens far away, such as a lightning strike, a flying plane, and a boom. However, it is hard for humans to estimate the range if the audio-visual event is closeby.  

Thanks to the advance in sensors, many mobile devices can support cameras with 240 FPS and even up to 960 FPS and microphones with 48 kHz sampling rate and up to 96 kHz). The high audio/video sampling rates make it possible to accurately measure the delay. The sensing scheme can leverage the audio-visual events to enable passive ranging. 

We observe that a single collision typically lasts within 16ms (\eg, 4 frames in 240 FPS) in the video. The impact sound is sometimes even shorter (\eg, 10 ms). Such a short collision makes it highly unlikely to have multiple collisions overlap in time. We analyze the video recording of the gym. We observe that impact sounds from different collisions almost do not overlap in time. In this paper, we do not explicitly handle overlapping collision sound, but our evaluation do include cases involving overlapping collision sound and show our scheme is fairly robust as long as the overlap is small. For more significant overlap, there are well-known sound source separation algorithms that can be used to separate these impact sounds. 

On the other hand, our system allows multiple objects that generate collisions to appear in the same video. This is achieved by first identifying all objects that cause collisions in the video and then applying the same method -- FBDepth to each object for range estimation.

%% file: approach2.tex
\section{System Design}
\label{sec:approach}


\begin{figure}[h!]
\centering
\includegraphics[width=\linewidth]{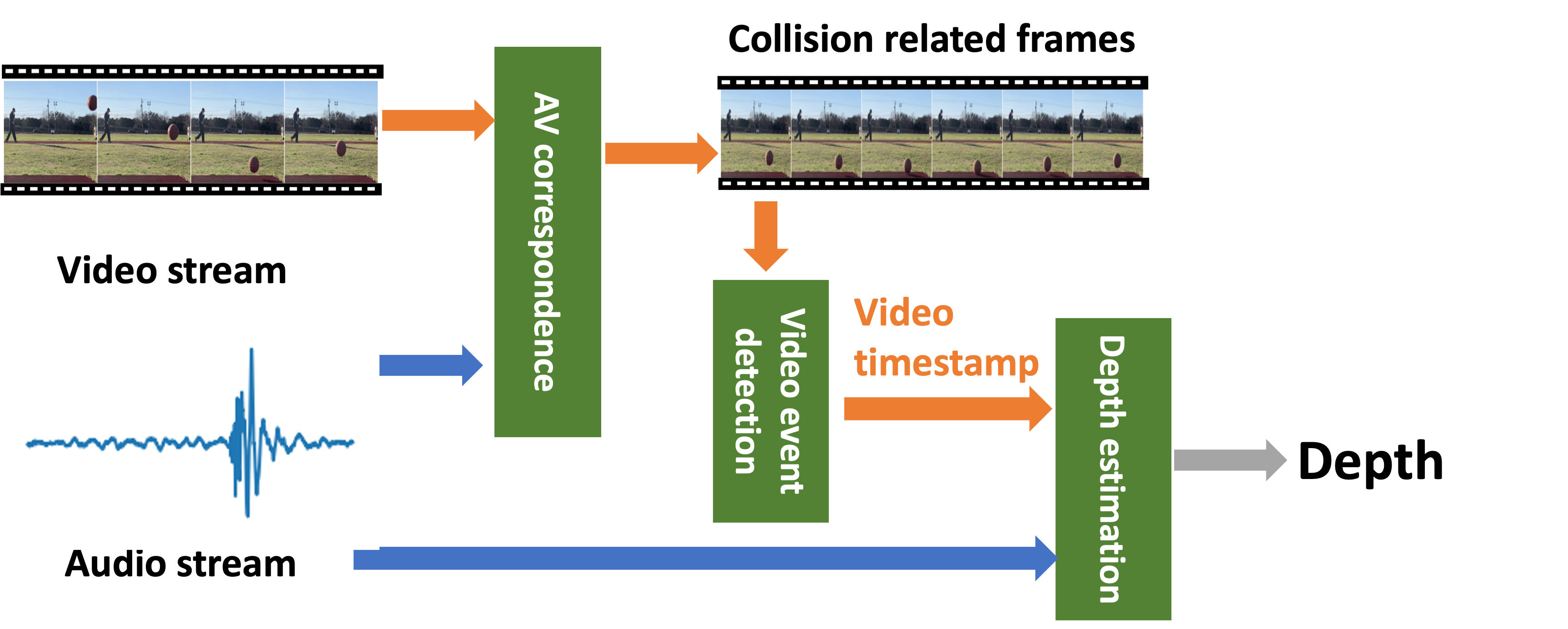}
\caption{DNN system.}
\label{fig:overview}
\end{figure}



As Figure~\ref{fig:overview} shows, we design a novel pipeline to identify clues from video and audio recordings and estimate the depth of the AV events. It consists of three important steps. 

First, it uses the audio-visual correspondence module to analyze the incoming audio and video streams to identify the video frames corresponding to the impact sound from the audio clip. This step detects the video and audio events at a coarse granularity. While the audio and video streams are not strictly synchronized, their offset is around 1 ms on iPhone, which is sufficient to support this coarse-grained event detection.  Furthermore, this step identifies the object in the video that produces the collision sound and generates the attention map to focus on the target. 

Second, it leverages the observation that a collision changes the direction of the motion to split video frames into two groups of frames: one before the collision and another after the collision. It interpolates the trajectories before and after the collision and uses the intersection to determine the collision time in the video $T_{video}$. 

Third, it further analyzes the audio recording to estimate $T_{audio}$ and derive $d$.  



\subsection{Audio-Visual correspondence}
\label{ssec:audio-event}


We first specify the audio-visual correspondence problem. Then we introduce optical flow. Next we present our motion aware AV correspondence network and algorithm based on the optical flow. We further describe how to support multiple collisions.  

\subsubsection{Problem Definition} 

Our first step is to detect the impact sound from the audio clip and identify the video frames that correspond to the impact sound. Since collisions are typically short (\eg, up to 16 ms), we just need to identify a few video frames that correspond to the impact sound. As the synchronization offset between the audio and video is small enough to support frame level event detection, we assume the audio and video streams are synchronized in this step. Given the incoming audio and video streams, our goal is to generate the following two outputs: (i) video frame classification and (ii) segmentation mask. For each video frame, we label it with 1 if the video frame corresponds to the impact sound and 0 otherwise. Moreover, for the video frame with label 1, we further generate segmentation mark where the pixels corresponding to the object involved in the collision are marked 1. The mask will be all zeros for video frames with classification label of 0. The segmentation mask highlights the colliding object and facilitates video event detection in stage 2. 

%


\subsubsection{Motion Aware Audio-Visual Correspondence Network}
\label{ssec:audio-visual-align}

\para{Primer of optical flow:}
\label{sssec:optical-flow}
The optical flow captures the movement of the brightness pattern, which is used to approximate the motion field.  More specifically, for each pixel $(x,y)$ in a video frame, the optical flow, denoted as $u(x,y)$ and $v(x,y)$, specifies the point's projection of motion in 2D. Assuming the "intensity constancy", we have 
$I(x+\Delta x, y+\Delta y, t+\Delta t) \approx I(x,y,t).$

Meanwhile, applying a Taylor series expansion, we get
$I(x+\Delta x, y+\Delta y, t+\Delta t) = I(x,y,t) + \frac{\delta I}{\delta x} \Delta x + \frac{\delta I}{\delta y} \Delta y + \frac{\delta I}{\delta t} \Delta t.$ Hence we arrive at 
$$ \Delta I v + I_t = 0$$
where $v = (\Delta x, \Delta y)$ is the optical flow and $\Delta I = (I_x, I_y)$ is the spatial gradient, and $I_t$ is the temporal gradient. By deriving $\Delta I$ and $I_t$ from a pair of video frames, we can obtain the optical flow $v$. 

Lucas-Kanade method is the most famous algorithm for computing the optical flow. It derives multiple constraints from the pixel along with its neighbors and uses the least square to estimate $v$. More recently, many deep learning algorithms have been developed to improve optical flow estimation. We choose to use the RAFT~\cite{RAFT}, which is one of the latest algorithms with excellent performance. It computes the optical flow in a multi-scale manner, where at each scale it uses a ConvGRU module to iteratively estimate the optical flow and move the pixel according to the derived flow.

\begin{figure}[h!]
\centering
\includegraphics[width=\linewidth]{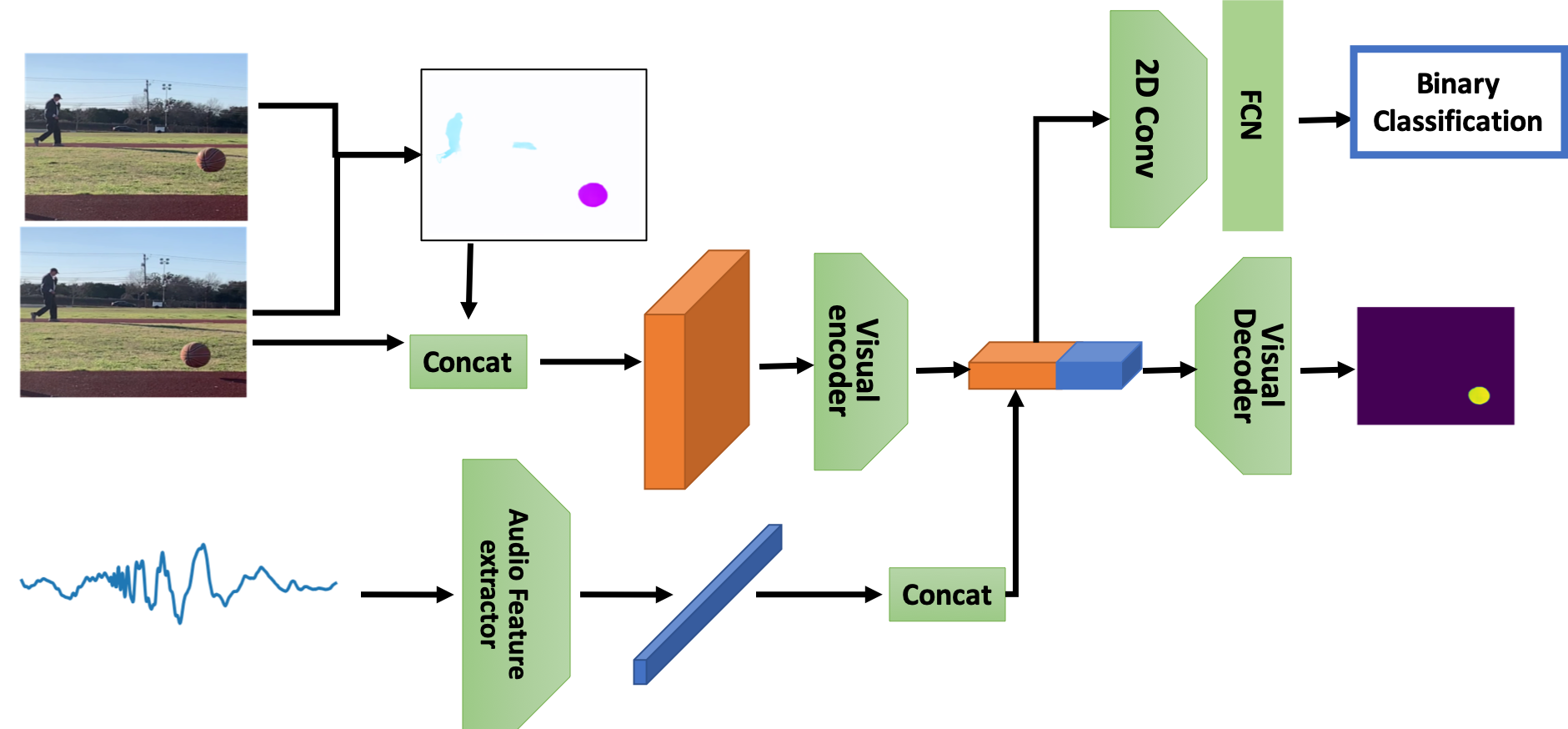}
\caption{MVANet for audio visual correspondence.}
\label{fig:nn-correspond}
\end{figure}

\begin{figure}[h!]
\centering
\includegraphics[width=0.8\linewidth]{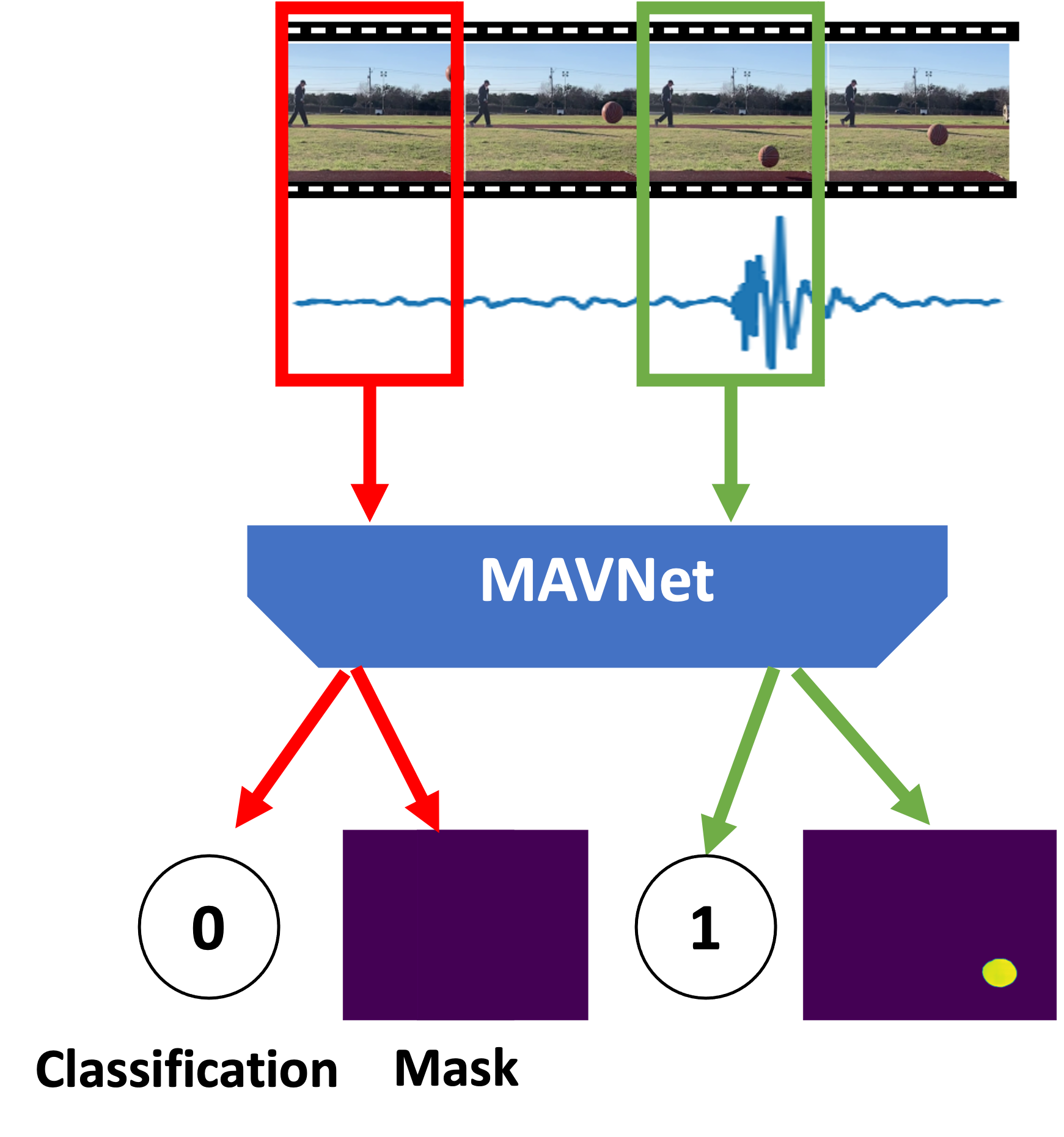}
\caption{Pair the audio visual stream by MVANet}
\label{fig:nn-pair}
\end{figure}


\para{Motion Aware Audio-Visual correspondence Network:}
\label{ssec:audio-visual-align}

Our goal is to detect impact sound and output video frame classification and segmentation masks according to the impact sound. We propose a motion-aware audio-visual correspondence network (MAVNet) for this purpose. MAVNet differs from the existing audio-visual learning networks in that (i) it jointly uses audio and video streams to extract the impact sound and (ii) it uses the optical flow as an additional feature so that it pays attention to the moving objects. 

One way to detect the impact sound is to use a threshold. However, this is not robust against background noise and interference. Instead we use audio and video streams together to detect the impact sound so that the impact sound in the audio corresponds to the moving target in the video. 

Meanwhile, MAVNet computes the optical flow between the two adjacent video frames. Let $I_t$ denote a video frame at time $t$. The optical flow $f_{t  \rightarrow t - \delta t}$ has the same width and height as frame $I_t$. It has two channels to capture the vertical and horizontal pixel-to-pixel movement from the current frame $I_t$ to the previous frame $I_{t - \delta t}$, respectively.  We concatenate $I_{t}$ with $f_{t \rightarrow t - \delta t}$ pixel by pixel to be a RGB-F tensor. Figure ~\ref{fig:nn-correspond} shows an example. The optical flow captures three moving objects: a basketball, human, and car. It filters out static pixels efficiently and provides a set of regions of interest. 

The RGB-F tensor is fed into a U-Net~\cite{unet}, which is a widely used network for image segmentation. It uses a series of convolution layers to extract visual features. Meanwhile, the audio clip is transformed by the audio feature extractor into a feature embedding. We replicate the audio feature, tile them to match the visual feature dimension, and concatenate the audio and visual feature maps. Afterwards, a series of up-convolutions are performed on the concatenated audio-visual feature map to generate a segmentation mask. Meanwhile, the audio-visual feature map is also fed into a small classification network with a series of convolution layers and linear layers to output a probability map for different AV events, where the probability indicates whether the current frame should have label 1 (\ie, corresponding to the impact sound).  

We use a combined Binary Cross Entropy (BCE) loss for both the classification and segmentation mask, which is defined as
 $$ Loss = BCE(mask, \hat{mask}) + BCE(y, \hat{y}) $$
 where $y$ and $mask$ refer to the AV event label and the segmentation mask, respectively. We label the classification manually and then generate and refine the segmentation mask based on the Siammask~\cite{wang2019fast} tracking network. 
 
\para{Model Implementation: } We use the pre-trained model from RAFT~\cite{RAFT} to compute the optical flow. This model is pre-trained on a mix of synthetic data and real data~\cite{2021mmflow}. It performs well on our current dataset. 
We pretrain a ResNet-18~\cite{ResNet} audio classifier on our own dataset. This model without the final layer will work as the audio feature extractor in the MVANet. It will output the audio feature embedding with a size of 256. The weights are also updated using MVANet. The U-Net has 5 down-convolution blocks to increase the number of channels to 16, 32, 64, 128, 256 and 5 up-convolution blocks to decrease the number of channels to 256, 128, 64, 32, 1.

{
\begin{algorithm2e}[t!]
$V, A = VideoStream(), AudioStream()$\\
\While{$V, A$ is available }{
    $I_0, I_1, t = V.prevFrame, V.curFrame, V.curTime$\\
    $f = RAFT(I_0, I_1)$\\
    $a = A(t)$\\
    $cls, mask = MVANet(I_1, f, a)$\\
    \If{cls is valid}{
        track the object forward and backward $a$ \\
        perform video event detection and depth estimation\\ 
    }
}

\caption{Audio-visual correspondence algorithm pseudo-code}
\label{fig:pseudocode}
\end{algorithm2e}
}

\para{Audio-visual stream correspondence algorithm:}
Based on the MVANet, we develop an audio-visual stream correspondence algorithm~\ref{fig:pseudocode} to align the incoming audio and video streams. We use the video with low FPS to decrease the computation cost. It is sufficient to detect the AV event with 30 FPS. Each frame lasts 33ms. Meanwhile, the duration of 2 frames can cover the impact sound. Therefore, we calculate the timestamp $T_{I_n}$ based on the frame count number and crop the audio samples in the range $[T_{I_{n-1}}, T_{I_{n+1}}]$. Then we use MVANet to check whether the audio clip from the impact sound can be paired with the two video frames in time. After we pair the audio and video events, we use the property of motion continuity to find the necessary frames to cover the whole collision process. Note that this pairing is not affected by $T_{hardware}$ since $T_{hardware}$ is around 1 ms on iPhone, which is too small to affect pairing. 






\para{Multiple colliding objects:} 
We can classify the cases involving multiple colliding objects based on whether the objects' visual appearance is similar and whether their impact sound is similar. 
As described in Section~\ref{ssec:av-spatial}, it is highly unlikely to have multiple collisions overlap in time. Therefore, in the following discussion, we mainly focus on identifying the video frames corresponding to the impact sound arising from one collision. We can simply apply the same procedure to each impact sound sequentially when there are multiple impact sounds.

\subsection{Video Event Detection}
\label{ssec:video-event}
Given a set of video frames including the whole collision process, we can visually locate the collision moment by observing the difference between two consecutive frames. The accuracy depends on the frame rate. It is 4ms for 240 FPS, which is insufficient for accurate depth estimation. Unfortunately, we cannot access the ground truth timestamp of the collision moment. It is challenging to design and tune a method without the ground truth. To address this challenge, we propose a coarse-grained to fine-grained estimation algorithm based on the motion consistency and change during the video event.


\subsubsection{Motion Consistency}
Motion consistency has been explored widely in video processing to help with spatio-temporal tasks. If the motion of the object is temporally stable across several frames (\eg, due to a constant force), the position and pose of this object can be predicted in the future frames as well as be interpolated between two frames. However, the impact sound is always caused by an impulse force, which results in a rapid change of the motion status. It breaks the motion continuity and consistency, and makes the interpolation around the time of collision more challenging. 

To handle the abrupt change, we seek a new form of motion consistency.  
Figure~\ref{fig:interp-example} shows a tennis ball is bouncing on the stone table. The numbers refer to the frame indices. As shown in the first three frames, the ball has a consistent downward and forward motion along with self-rotation before the collision. Then it bounces up after touching the surface. The direction and magnitude of the velocity change rapidly by the elastic force. 
We want to estimate the collision time between the frames $I_3$ and $I_4$. Traditional video frame interpolation (VFI) uses the motion consistency to interpolate the frames between $I_2$ and $I_3$. However, VFI cannot interpolate frames correctly between $I_3$ and $I_4$ due to an abrupt change between the two frames. In fact, it is ambiguous to tell the position between $I_3$ and $I_4$ by observing these two frames. 

Fortunately, this collision also has motion consistency. First, the motions before and after the collision are consistent individually. Therefore, we can interpolate and extrapolate the position of the ball using points before the collision (\eg, $I_1, I_2, I_3$). Similarly, we can also estimate the trajectory using the points after the collision (\eg, $I_4, I_5, I_6$). 
Second, the intersection between these two trajectory is the collision point. Therefore, we can estimate the collision time by finding their intersection. 

\subsubsection{Coarse estimation of collision time} 

Given a sequence of video frames, 
we first split them into two sets: the frames before the collision $\mathbf{V_0}$ and the frames after the collision $\mathbf{V_1}$. This essentially requires us to determine the last frame $I_{e}$ in $\mathbf{V_0}$ before the collision and the first frame $I_{s}$ in $\mathbf{V_1}$ after the collision. 

Based on the analysis of the physical motion, we make an important observation that can help us determine $I_{e}$ and $I_{s}$. The collision results in a significant acceleration change due to the strong impulse force. Let $a_t = v_t - v_{t-1}$ and $\delta a_t = a_t - a_{t-1}$ denote the acceleration and acceleration change of frame $I_t$, respectively. $\delta a$ between $I_{e}$ and $I_{s}$ is large, while $\delta a$ between adjacent frames before the collision (or after the collision) is small. 
If the object stops moving immediately after the collision, we take the static frame $I_{e+1}$ as $I_{s}$. Finally, we select the frames before $I_{e}$ to generate $\mathbf{V_0}$, and select the frames after $I_{s}$ to generate $\mathbf{V_1}$. 

We use the efficient single object tracker to determine the object positions in the frames and calculate the velocity, acceleration, and acceleration change. The estimation of $I_{e}$ gives a frame-level  resolution (\eg, 33ms for 30 FPS or 4.2 ms for 240 FPS).

\subsubsection{Fine-grained estimation of collision time} 
\label{sssec:interp}

\begin{figure}[h!]
\centering
\includegraphics[width=0.8\columnwidth]{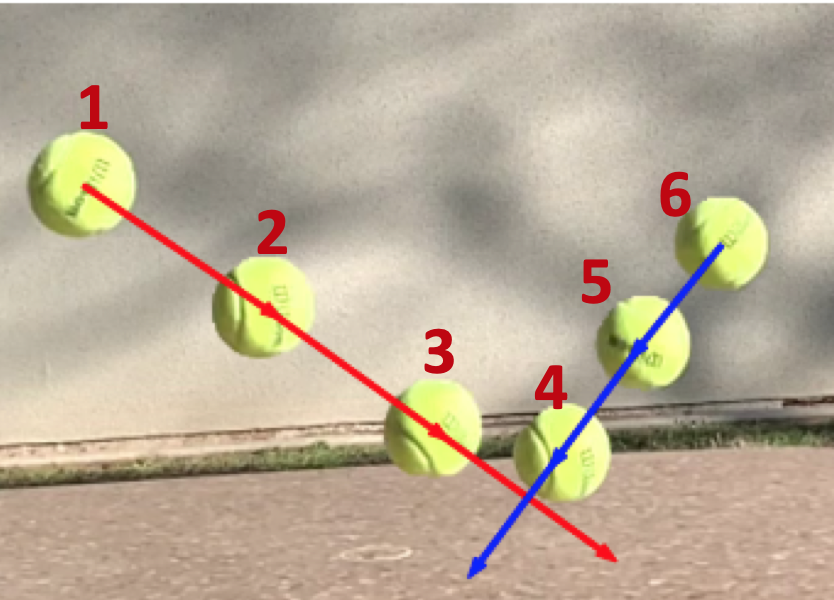}
\caption{Interpolate trajectories before and after collision to compute the intersection for collision detection.}
\label{fig:interp-example}
\end{figure}

Considering the $1ms$ temporal error results in a $35cm$ depth error, the frame level resolution is not sufficient for the depth estimation. We can extrapolate the trajectories of $\mathbf{V_0}$ and $\mathbf{V_1}$ to locate the same collision moment. A simple solution is to use the center of the ball to fit the 2D moving trajectory, and use the intersection of two trajectories as the collision point. However, this method is not accurate for real experiments. The center point simplifies the shape and color of the object. It loses rich information of thousands of pixels from the object. For the example in Figure ~\ref{fig:interp-example}, the center point cannot model the pose or rotation of the object. Thus, it does not take advantage of all visual constraints. Moreover, the deviation of each center's position can lead to a considerable error because the data points are not sufficient.

We apply optical flow to enable a fine-grained robust video event estimation. Optical flow can provide pixel-to-pixel motion in the 2D frame projected from the 3D space. We can use all pixels to search for the optimal collision time. \Note{add a figure to show the difference} The optical flow sequence is usually generated by adjacent video frames. Considering a set of frames $\{I_0, I_1, ...I_n\}$, we can estimate a flow sequence $\{f_{0 \rightarrow 1}, f_{1 \rightarrow 2}, ... f_{n-1 \rightarrow n}\}$.  For a pixel $p_0(x, y)$ of the target object in the frame $I_0$, we can infer the new position of the pixel in the frame $I_1$, denoted as $p_0(x+\Delta x_1, y + \Delta y_1)$, where $(\Delta x_1,  \Delta y_1) = f_{0 \rightarrow 1}(x, y)$. Based on the optical flow sequence, we can estimate the coordinate of this pixel in the next few frames as $$\{(x, y), (x+\Delta x_1, y + \Delta y_1), ..., (x+\sum_{i=1}^{n} \Delta x_i, y + \sum_{i=1}^{n} \Delta y_i)\}$$.


However, this method will result in error accumulation from the optical flow and mismatch the pixels across the frames. Therefore, we compute the optical flow from an anchor frame $I_a$ to all other frames. We select the last frame $I_e$ from the set $\mathbf{V_0}$. It is near the collision moment, so its optical flow to the other frames around the collision time is relatively smooth and easier to estimate. Let $\{I_0, I_1, ...I_n\}$ denote a set of frames. We can estimate the flow sequence as $\{f_{e \rightarrow 0}, f_{e \rightarrow 1}, ... f_{e \rightarrow n}\}$. For a pixel $p_e(x, y)$ of the target object in the frame $I_e$, we can infer the relative position of the pixel in the frame $I_0$, denoted as $p_{e \rightarrow 0}(\Delta x_0, \Delta y_0)$, where $(\Delta x_0,  \Delta y_0) = f_{e \rightarrow 0}(x, y)$. The coordinate sequence can be represented as
$$\{(\Delta x_0,  \Delta y_0), (\Delta x_1,  \Delta y_1), ... ,(\Delta x_n,  \Delta y_n)\}$$
Therefore, we avoid the error accumulation. We select $k$ consecutive frames as subsets $\{I_e, I_{e-1},...,I_{e-k+1}\}$ and $\{I_s, I_{s+1},...,I_{s+k-1}\}$ from $\mathbf{V_0}$ and $\mathbf{V_1}$ respectively. We are interested in the optical flow of the target object, which is obtained by the optical flow of the entire frame filtered using the segmentation mask. It includes hundreds to thousands of pixels. 

For each frame $i$ in $I$, we compute the optical flow between frame $i$ and anchor frame $a$, denoted as $OF_{i,a}$. Then we perform curve fitting on $OF_{i,a}$. We vary the size of $I$ and the types of curves for interpolation. We find using a linear fit and $|I| = 6$, which means 3 video frames before and after the collision each yields the best performance. 

We illustrate an example in Figure~\ref{fig:interp-example}. We compute the optical flow between the pixels in $I_i$ versus $I_a$, where frame 3 is the anchor frame. The red and blue curves show our interpolated trajectories based on the optical flow of the frames before and after the collision, respectively. Since we find a linear fit yields the best performance, we determine the collision time based on the intersection between the two lines corresponding to the trajectories before and after the collision.

\begin{figure}[h!]
\centering
\includegraphics[width=\columnwidth]{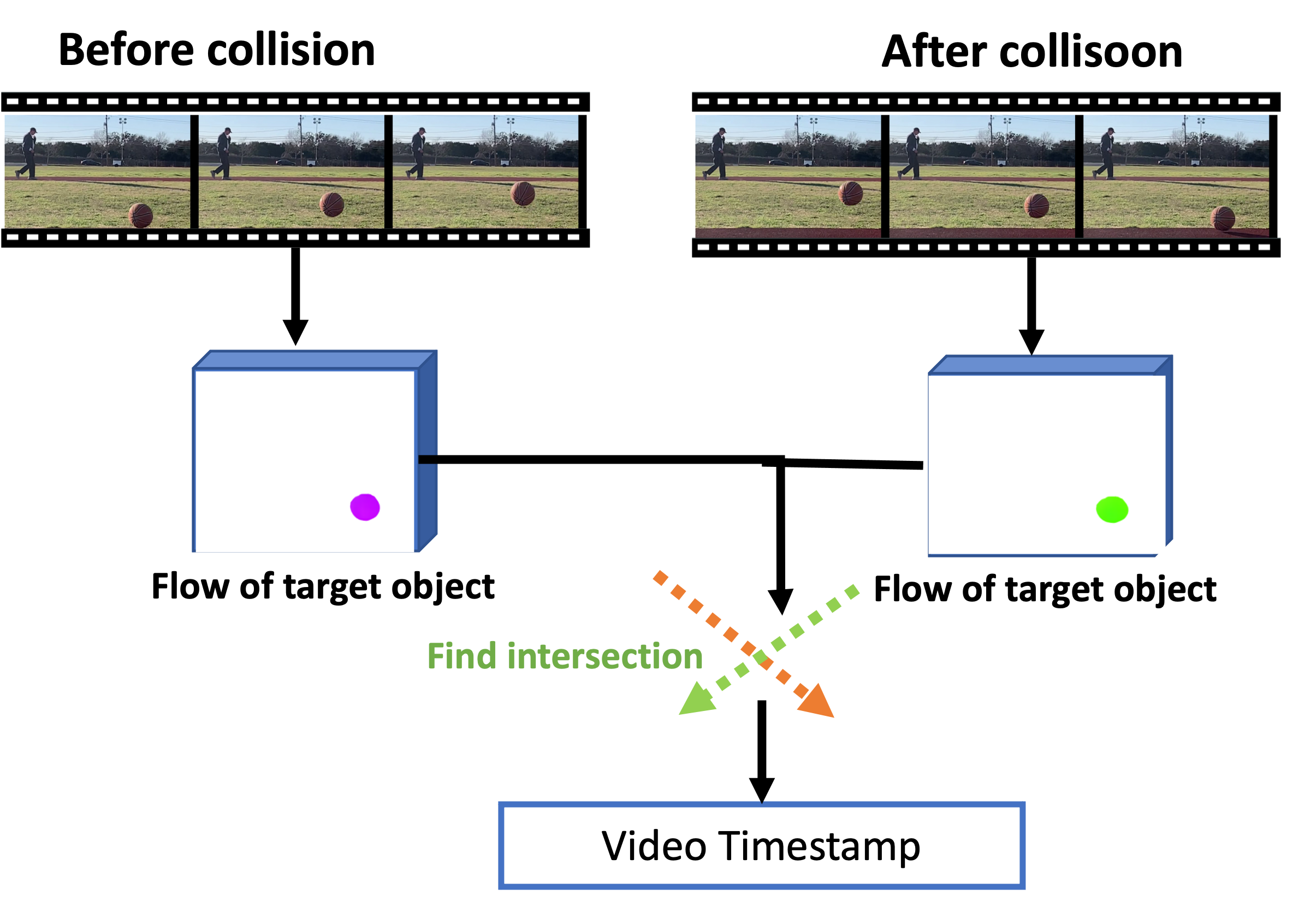}
\caption{Find estimation from the collision frames.}
\label{fig:interp-find}
\end{figure}

\subsubsection{Collision Time Estimation}
\label{sssec:collision-est}
\Wei{even current formulation is not accurate}
Note that a target has many pixels, each with an optical flow estimate. Let $P$ denote the set of pixels corresponding to the target. For each pixel $p$ (\ie, $p \in P$), we get a series of optical flow during its movement before and after the collision and interpolate and compute their intersections as described above. Let $I_p$ denote the intersection between the pixel $p$'s before and after trajectories, and $I_p(t)$ denote the time at which the trajectories before and after the collision intersect. To determine the collision time for the target, we take advantage of all pixels' intersection points by finding the time $t$ that minimizes the intersection loss, namely $min_t\ \sum_p |t - I_p(t)|$. 

\begin{figure*}[h!]
\centering
\includegraphics[width=1.8\columnwidth]{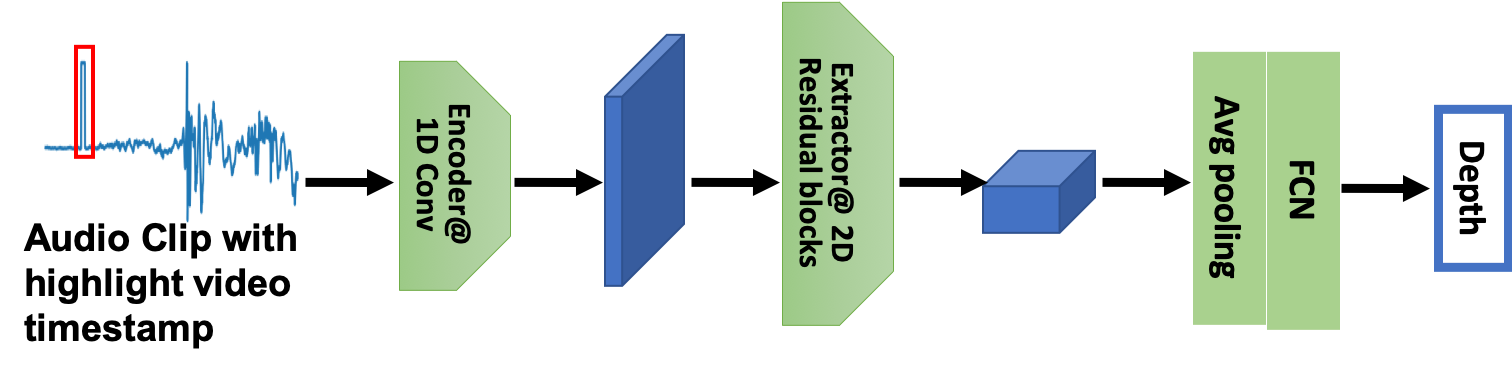}
\caption{DNN to detect the collision from the audio recording and estimate the distance.}
\label{fig:nn-audio}
\end{figure*}

\subsection{Distance Estimation}
\label{ssec:dist-est}
\Wei{in the real environment, the sound generation procedure varies a lot across different objects ans materials; the collision moment in the visual can still has a offset to the audio generation; describe the wave first, early wave is small, later peak is large; cannot distinguish by human; }
By now we have obtained the collision timestamp in the video with a millisecond time resolution. Even though the sampling rate is sufficient in the audio samples, it is still challenging to determine which samples correspond to the collision moment as shown in Figure~\ref{fig:audio-demo}, where some objects (\eg, wooden bricks) generate strong impulses immediately upon collisions  while others (\eg, balls) generate weak sound followed by strong sound. Therefore, manually labeling the collision time in the audio recording is not only time consuming but also unreliable. Another important problems is the variable delay between the camera and microphone. The timestamp of a camera on iPhone is at a microsecond resolution, but its microphone has more variability.


To bypass the manual labeling of collisions in the audio recording, we do not attempt to detect the collision event in the audio recording explicitly and use the difference between the collision event in the audio and video recordings to derive the distance. Instead, we transform the video timestamp into audio sample index and highlight the surrounding audio samples by marking them as $1$. We develop a depth estimation model to automatically regress the depth with the highlighted pattern and waveform. 

 It takes a short sequence of audio samples (\eg, 1600 samples) close to the collision time detected in the video as the input, where the audio samples very close to the video collision time (\eg, 24 samples) are highlighted. As shown in Figure~\ref{fig:nn-audio}, the peak in the red rectangle is the highlighted area, which conveys the video timestamp. The 1D convolutional encoder uses a small window (\eg, 16 samples)  and a small stride (\eg, 8 samples) to encode temporal samples into embeddings. Compared with Short Time Fourier Transform (STFT), which is a traditional encoder with a much lager window size and stride size, we use a small window to achieve a good temporal resolution. Next we have 2D residual blocks to identify the delay pattern hidden in the feature map. Finally, a fully connected layer maps the embeddings from the pooling layer to the estimated depth. We use Mean Square Error (MSE) between the estimated depth $\hat{d}$ and ground truth depth $d$ (\ie, $L(d, \hat{d}) = mse(d, \hat{d})$) as the loss function.

%% file: eval.tex
\section{Implementation}
\label{sec:eval}
In this section, we describe how to collect the video dataset and prepare the training set. 

\subsection{Dataset Collection}
\begin{figure}[h!]
\centering
\subfigure[Data collection platform]
{\includegraphics[width=0.48\columnwidth]{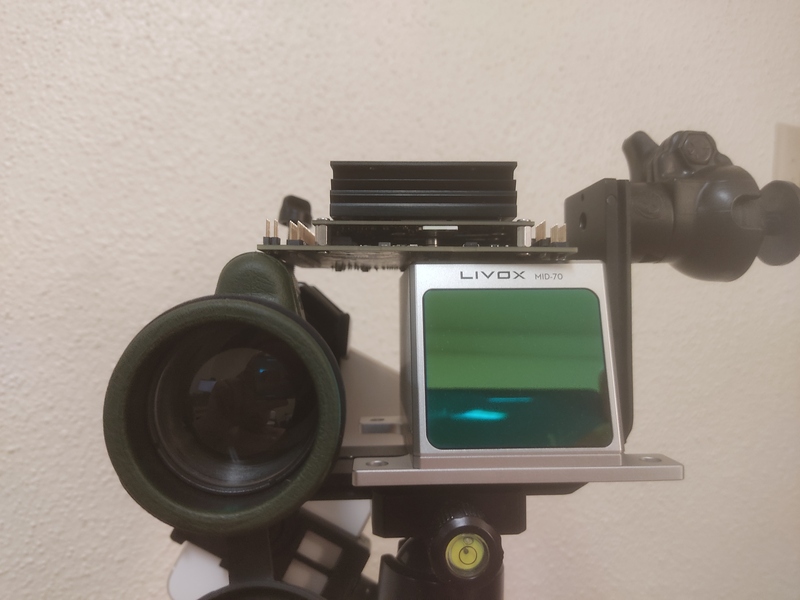}}
\subfigure[Experiment objects]{\includegraphics[width=0.48\columnwidth]{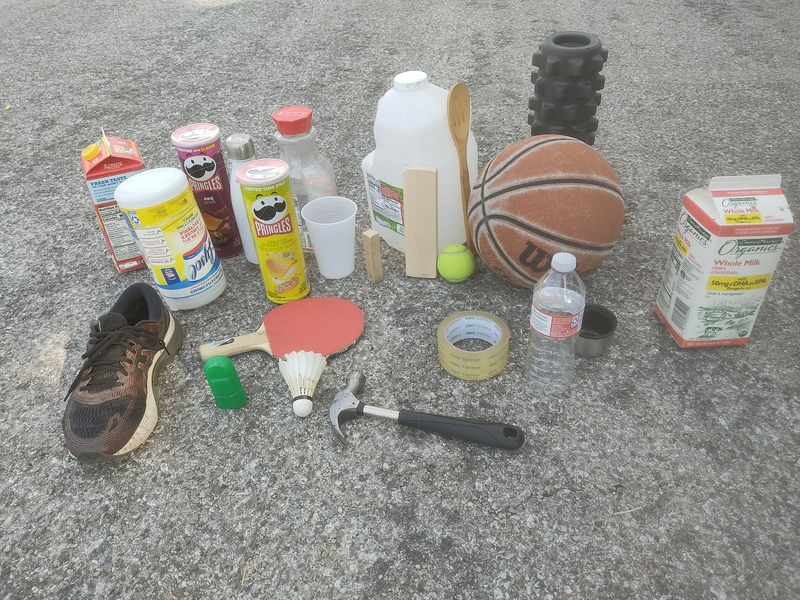}}
\caption{Data collection platform and objects }
\label{fig:hard}
\end{figure}

Figure~\ref{fig:hard} shows our data collection platform. We use an iPhone XR with a slow-motion mode to collect the video data and audio data. The iPhone's video recording and audio recording are synchronized well at low-level API.  The timestamp differs by a small delay ($~$ 1 ms) with a small variance (within 1 ms). Moreover, its frame rate is stable and does not have a frame rate drifting. Thus, we can transform the frame number to the timestamp accurately. Besides, we mount an ARPBEST monocular telescope to the phone to improve the vision at a large distance. In fact, a telescope camera has become an indispensable camera in recent years. Samsung Ultra 20 can support 10x optical zoom and 100x hybrid zoom, and Pixel 6 pro has 20x zoom. 
We compare the photos taken by iPhone with the extra telescope with the photo taken by Pixel 6 pro. We observed that the image quality of our setup is a bit worse than the Pixel's image. Overall, our setup resembles the hardware available on commercial mobile phones. 

To get the ground truth depth for quantifying the estimation error of other methods, we use a LIVOX MID-70 LiDAR. It emits the non-repetitive scanning pattern, collects LiDAR points for a few seconds, and merges all the points into one frame to get a dense point cloud. This setup is appropriate for scanning the static scene. However, it does not work for moving objects. 
To solve this problem, we select a static anchor at the target depth and use the LiDAR to measure the depth to the anchor. Then we perform collision experiments at the position of the anchor. The actual collision point may deviate from the anchor position, but the error is negligible compared to the overall estimation error. 

We compare our approach with monocular camera based solution. We find that the latest monocular depth estimations cannot adapt to our dataset well because the scenes in our dataset are quite different from those in the training data. Therefore, we use Monodepth2~\cite{godard2019digging} with its pretrained model and fine-tune the model using our data. We collect dense point clouds of static scenes to generate depth maps to help fine-tune Monodepth2. 

Figure ~\ref{fig:hard} shows 23 objects used to produce the impact sound. They include various masses, sizes, and shapes. These objects cover six common materials: wood, metal, foam, rubber, plastic, and paper. We drop objects to the floor to make the collision. These are regarded as the free-fall motion. In addition, we also include two types of human motion: walking and striking the hammer. These motions are affected by the force of the human as well. We collect the collision event from 2 meters to 50 meters in different environments, including the apartment, basketball court, office building, campus pavement, and residential areas. 

After data cleaning and annotation, we get around 3 thousand audio-visual sequences., including 210K+ frames. Each sequence has around 40 ~120 frames and an audio clip corresponding to these frames. The raw sequence are in 240 FPS and can be downsampled to lower FPS. The audio sampling rate is 48Khz. 

\subsection{Training/Testing Generation} 

We use 80\% audio-visual sequences for training and the remaining samples for testing. All the following data generation process is in the closed loop of each dataset.

\para{Dataset for the audio-visual correspondence network: }  Even though there are 3K audio-visual sequences, we can generate many more valid and invalid audio-visual pairs. We segment an audio clip $A_{impact}$ of 66.7ms including the impact sound for each sequence. We sample 20 frames from each sequence and pair them with the audio clip $A_{impact}$ to generate the valid pairs. Invalid audio-visual pairs consist of two sources.  First, each sequence has plenty of audio clips without impact sounds. They are paired with the sampled frames as invalid pairs. Besides, we also use the  audio clip $A_{impact}$ from one sequence to the frames from another sequence that do not include the same object. Each sequence generates 20 invalid pairs to keep the data balance between valid and invalid pairs. Finally, we generate around 96K audio-visual pairs in the training set and  24K  pairs in the testing set.

\begin{figure*}[h!]
\centering
\begin{minipage}{0.6\textwidth}
\centering
\subfigure[Absolute distance error]{\includegraphics[width=0.48\columnwidth]{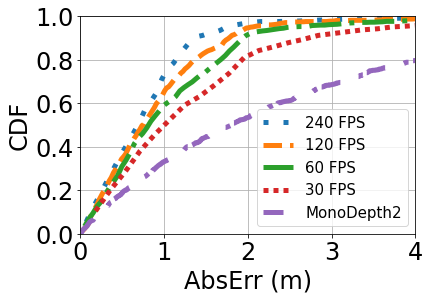}}
\subfigure[Absolute relative error]{\includegraphics[width=0.48\columnwidth]{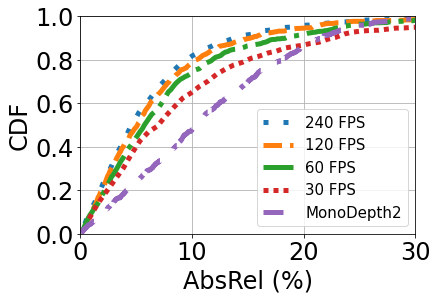}}
\caption{Depth estimation error at different frame rates. Compare the performance to the Monodepth2}
\label{fig:vary-fps}
\end{minipage}
\begin{minipage}{0.38\textwidth}
\centering
\includegraphics[width=0.96\linewidth, height=0.19\textheight]{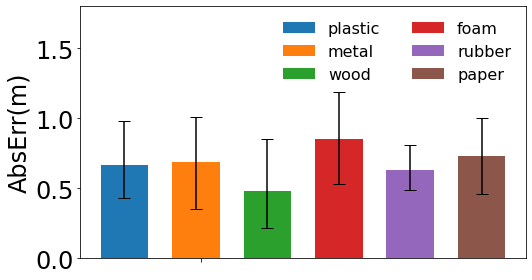}
\caption{AbsErr for various materials }
\label{fig:vary-mat}
\end{minipage}
\end{figure*}

\para{Synthesize a multiple collision dataset: } It is difficult to collect multiple collisions in the real environment. Therefore, we synthesize a video 
involving multiple collisions by adding collisions from other videos to the base video sequence that involves one collision. Specifically, we select one audio-visual sequence as the base sequence. We crop the objects from frames in another video sequence and patch them into frames at a random time $T_{patch}$ in the base sequence. We also copy the audio clip $A_{impact}$ in another sequence based on the timestamp $T_{patch}$. Based on the time of these collisoins in the merged sequence, we determine if the collisions overlap. If there is no overlap, we overwrite the audio segment using the audio segment corresponding to the newly added collision. Otherwise, we add the audio segment corresponding to the newly added collision to the original audio segment. 
Note that the cropped object and the cropped audio clip still hold the temporal relationship in the raw sequence. By now, we synthesize a dataset including 2 or 3 collisions. We use this dataset to generate a small dataset of audio-visual pairs and augment the single collision pairs. We generate around 25k audio-visual multiple collision  pairs in the training set and  6K  pairs in the testing set. 

\para{Depth estimation dataset: } It is identical to the augmented audio-visual sequence dataset. Each sequence will be applied by the audio-visual correspondence algorithm and the video event detection to generate the estimated video timestamp. The depth estimation dataset consists of these timestamps, audio clips, and ground truth depths. The training set has 3k samples while the testing set has 0.8K samples.  

\subsection{Model Implementation}
MAVNet and depth regression network are all implemented with Python and PyTorch. They are trained on the platform with an Intel(R) Xeon(R) Gold 6230 CPU @ 2.10GHz CPU
and an NVIDIA Quadro RTX 6000 GPU. The learning rate and batch size are set as 0.001 and 32 respectively. The benchmark Monodepth2 uses the pre-trained model on Kitti. It is finetuned on the collected depth maps to fit our data space with a learning rate of 1e-4.

\section{Performance Results}
In this section, we discuss the performance of FBDepth on the collected data. 
\subsection{End to end performance}

\para{Metrics} We use the absolute depth errors as $AbsErr = |D - \hat{D}|_1$ and absolute relative errors $AbsRel = \frac{AbsErr}{D}$ as the end-to-end performance metrics, 
AbsRel is the difference between the estimated and ground-truth distances $D$ and the estimated distance $\hat{D}$. AbsRel is the ratio between the AbsErr and the ground truth distance. It reflects the significance of the error to the ground truth. We vary the type of FPS, objects, distances, and environments to understand their impacts. 



\para{Overall Performance: } Figure~\ref{fig:vary-fps} shows the overall performance of FBDepth with different frame rate and Monodepth2. 240 FPS has the input of the best temporal resolution.  It can achieve a median AbsErr is 0.69m and a median AbsRel is 4.7\%. Although other frame rates have a lower resolution to capture the movement, FBdepth can still achieve promising performance. The median AbsErr is 0.99 m and the median AbsRel is 6.9\% for 30 FPS. Both metrics outperform Monodepth2. In comparison,  Monodepth2 can only achieve a median AbsErr of 1.80 m and a median AbsRel of 10.4\%. Moreover, the 75-percentile error in Monodepth2 is much larger than that of our method: 1.01 vs. 3.46m. In all, FBDepth benefits from the spatial propagation fundamental and outperforms monocular depth estimation which relays on the implicit depth structure and hints.  
\begin{figure*}[h!]
\centering
\begin{minipage}{0.7\textwidth}
\centering
\subfigure[Absolute distance error]{\includegraphics[width=0.49\columnwidth]{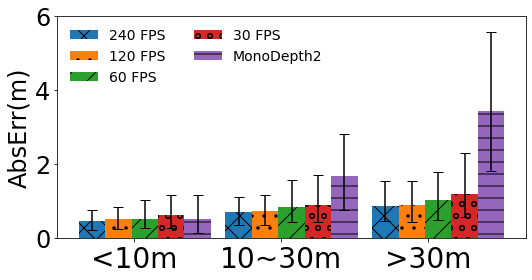}}
\subfigure[Absolute relative error]{\includegraphics[width=0.49\columnwidth]{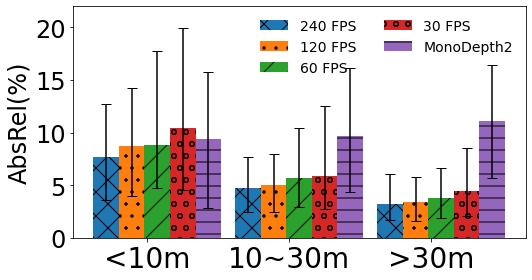}}
\caption{Depth estimation across different distances}
\label{fig:vary-distance}
\end{minipage}
\begin{minipage}{0.29\textwidth}
\centering
\includegraphics[width=0.96\linewidth, height=0.17\textheight]{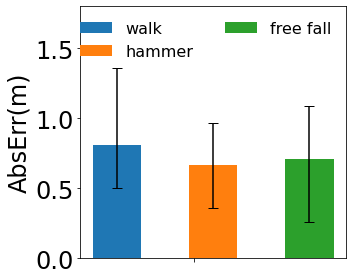}
\caption{AbsErr for various motions }
\label{fig:walk}
\end{minipage}
\end{figure*}

\para{Impact of frame rates:} Figure~\ref{fig:vary-fps} plots the absolute error and relative error  under different frame rates. 

As we would expect, increasing the frame rate reduces the estimation error of FBDepth. Increasing 30 FPS to 60 FPS yields the largest improvement, and the benefit gradually tapers off with a further increase in the frame rate. We observe that 30 FPS is too slow to capture sudden movements while 60 FPS is at the borderline. It is consistent with the trend to set 60 FPS as the default video recording and playing. 

\para{Impact of the distance:} Figure~\ref{fig:vary-distance} plots the AbsErr and AbsRel at 3 different range.
We group the ground truth distance into small (\ie, within 10 m), medium (\ie, 10-30 m), and large distances (\ie, farther than 30 m). 
Bottom, and top of the errorbars in the barplot denote median, 25-percentile, and 75-percentile, respectively. It is clear that FBDepth has a different trend compared to Monodepth2 in the AbsRel. While AbsRel does not change much in the estimation of Monodepth2, AbsRel decreases by the distance for FBDepth. Therefore, the AbsErr increases proportional to the distance for Monodepth2. Meanwhile, AbsErr of FbDepth increases slowly by the distance. 

For example, under 240 FPS the mean error is 0.45m at a short distance, grows to 0.70m at a medium distance, and grows to 0.85m at a large distance. The AbsRel drops to 3.2\% at the large distance.  This result shows the advantage of FBDepth since it is based on the physical delay and is not affected by the actual depth directly. Although the distance makes the number of effective pixels smaller and decreases the SNR of the impact sound,  they do not degrade the performance severely. Meanwhile, MonoDepth2 can only leverage  monocular depth hints to generate a good disparity map, which is not sufficient to the accurate value in the real world.

\para{Impact of object materials} We use the 23 objects as shown in Figure~\ref{fig:hard} 
These objects have many common characteristics and unique features. Among them, we observe that materials can be one of the important features. The material can affect both the frequency and magnitude of the impact sound. Meanwhile, it is related to the elastic force directly and can be important to the motion after the collision. The objects include six common materials. ~\ref{fig:vary-mat} shows the performance of each material. 

Among them, wood has the best performance. Its median AbsErr is 0.48m. We realize that wood always produces a loud sound and does not spin fast. The collision procedure of wood is smooth. Foam has the largesst median AbsErr of 0.85m. The impact sound is light and the object is not rigid enough. the other materials have similar performance.  Metal has a strong impact sound but it may spin fast. Rubber takes a lot of time to bounce but the sound is weak without a large velocity. 



\para{Impact of object motions} We can classify the object movement into free fall and human-forced motions.
Compared to the free fall driven by gravity, human-forced motions are more complex to predict. Fortunately, human-forced motions are consistent as well because it is rare for humans to perform a sudden force. We take two typical motions, walking and striking the hammer. As shown in Figure ~\ref{fig:walk}, walking has the largest median AbsErr as 0.81m. Compared to other methods, walking is a bit more challenging due to its small impact sound and a bit of deformation of shoes. Striking hammer can get an average result. It does not spin at all. Meanwhile, the impact is very strong. However, the human may change the force of hamming before and after the collision, which can bring more non-linearity to the model.  





\subsection{Micro Benchmarks}

In this section, we perform micro benchmarks to evaluate the benefit of each component in FBDepth. 



\begin{figure*}[h!]
\centering
\begin{minipage}{0.32\textwidth}
\centering
\includegraphics[width=0.98\linewidth, height=0.19\textheight]{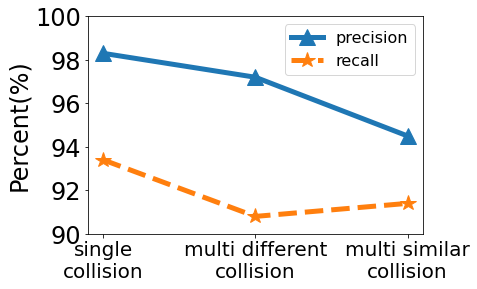}
\caption{Precision and recall for audio-visual correspondence in different collision setups }
\label{fig:micro1}
\end{minipage}
\begin{minipage}{0.67\textwidth}
\centering
\subfigure[Temporal error of the estimation of low FPS compared to 240 FPS]{\includegraphics[width=0.49\columnwidth]{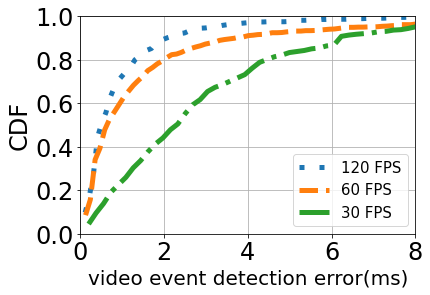}}
\subfigure[Improvement ratio of temporal resolution]{\includegraphics[width=0.49\columnwidth]{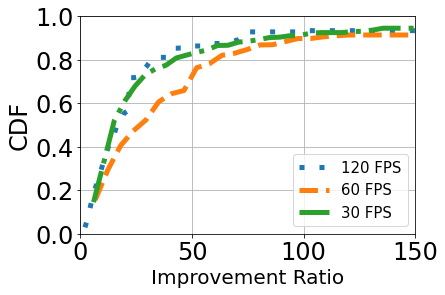}}
\caption{Effectiveness the video event detection in the second stage}
\label{fig:micro2}
\end{minipage}
\end{figure*}


\para{Impact of audio-visual correspondence:}
The first stage of audio-visual correspondence is able to identify the valid audio-visual pairs and transfer the single pair for further process. It is the key step to decoupling different objects and sounds existing in the video and audio. Figure ~\ref{fig:micro1} shows the performance of our audio-visual correspondence algorithm. A high precision stands for only a few invalid pairs in the selected pairs. A high recall reveals that most valid pairs have been discovered. When there is a single collision, both the precision and recall are the highest among the three cases. The recall is 4.9 \% smaller than the precision since there are many pairs with imperceptible sound. When there are multiple different collisions, the recall and collision drop slightly due to a small probability of overlapping collisions. The  degradation is small enough for further 
estimation. The most challenging case is multiple similar collisions. It degrades the precision to 94.5\% because it is challenging to identify the  correct pairs in some corner cases. The recall is slightly better as more pairs can be recognized. Once the audio and video events are correctly paired, the distance estimation errors under multiple collisions are similar to the errors under a single collision case. 

\para{Robustness of video event detection}
There is no ground truth of the video event so we cannot directly quantify the accuracy of video event detection. However, we propose a method to validate the coarse-to-fine algorithm. We set the estimation of 240 FPS as the baseline. We compare the estimation of other lower FPS. If these results cannot converge to similar numerical results, then we can decide the algorithm is not accurate and robust. Otherwise, it can get similar results from independent input, which means the algorithm is reliable. 

We show the temporal error of 30, 60, and 120 FPS compared to  240 FPS in Figure \~ref{micro2}. The median error for 30, 60, and 120 FPS is 2.3ms, 0.65ms, and 0.5ms respectively. It reveals that all the estimation results are much closer to the baseline of 240 FPS. Considering the frame resolution, we can compute the improvement ratio as $\frac{frameDuration}{temporal_error}$
For example, if the frame duration of 30 FPS is 33.3 ms, we can get the median improvement ratio as 14.5. The 60 FPS has the largest 25x improvement. This is a strong evidence that our coarse-to-fine estimation algorithm is reasonable and robust. 



\begin{figure}[h!]
\centering
\includegraphics[width=0.8\columnwidth]{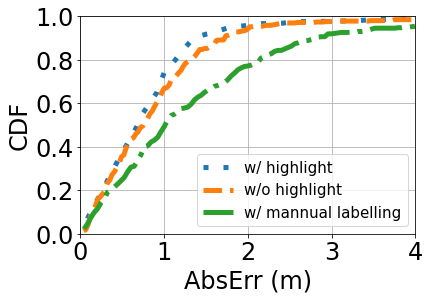}
\caption{Impact of direct versus indirect audio learning.}
\label{fig:eval-dir}
\end{figure}

\para{Impact of direct vs indirect audio learning:} In step (iii), we train a DNN to directly estimate the depth based on the audio samples. Alternatively, we could train a DNN to detect the collision in the audio recording. But this requires one manually label the collision time in the audio, which is both time-consuming and error-prone. We compare these two approaches in Figure~\ref{fig:eval-dir}. Even if we take the time to manually label the collision in the audio, its performance degrades due to labeling error: its median error increases from 0.69m to 1.01m. This demonstrates the effectiveness of our direct audio learning in step (iii).  Besides, we also try not to highlight the video timestamp into the audio clip (e.g. w/o highlight). We directly train the audio clip and take the difference to the video timestamp. The median error is 0.75m.

%% file: related.tex
\section{Related Work}
\label{sec:related}

We classify the existing work into the following categories: (i) depth estimation, (ii) video prediction and interpolation, and (iii) audio-visual learning.

\para{Depth estimation:} As mentioned in Section~\ref{sec:intro}, there have been many types of depth estimation methods. Some require special hardware, such as depth sensors and Lidar, which are not widely available on mobile devices yet. RADAR technologies using different radio frequencies or acoustic have limited ranges or face multipath challenges. Camera based approaches are also popular. Some require multiple cameras, a single moving camera, or dual pixel hardware to get different perspectives. Others are based on a single camera and more widely applicable. Refer to \cite{monocular1,monocular2} for detailed surveys on monocular depth estimation. They leverage feature learning, cost volume estimation and regularization, disparity estimation, post-processing and refinement, or learning confidence and uncertainty. 

A few recent works explore sound source localization~\cite{sslactive,ssllocalization,sslsurveillance,refaware,ssl-nsdi21}.  \cite{refaware} uses a Kinect depth sensor to construct a 3D model of the environment and then measures the AoAs of the multiple paths the sound source traverses and uses re-tracing to localize the source. Its localization error is $1.12m$. \cite{diffaware} considers the diffraction paths in addition to the reflection paths and achieves an error of $0.82$ m. 
\cite{ssl-nsdi21} uses inaudible audio signals to reconstruct the room structure without special hardware and improves the AoA estimation by explicitly considering the coherence in the multipath. It achieves 0.31 m error in line-of-sight (LoS) and 0.47 m in non-line-of-sight (NLoS).

\para{Video prediction and interpolation:} Video prediction has been a popular research topic due to many applications, such as video recovery, autonomous driving, and robot navigation. \cite{video-pred-survey} provides a nice survey of video prediction. Various models have been proposed, including convolutional networks (\eg, \cite{cnn1,cnn2,cnn3,cnn4}), recurrent networks (\eg, \cite{rnn1,rnn2}), generative networks (\eg, density based~\cite{density1,density2,density3}, and auto-encoder based \cite{autoencoder1,autoencoder2}). Designing an appropriate loss function has a significant impact on the video prediction accuracy. Due to the uncertainty of future frames, minimizing conventional loss functions, such as MSE and $l_2$ losses, would result in averaging the prediction across multiple possibilities and yield blurred images. Perception losses (\eg, \cite{loss2,loss3}) have been proposed to mitigate the issues. GAN has been widely used for video prediction. It consists of a generator and discriminator, where the discriminator tries to discriminate between the real and fake images and the generator tries to generate images that are deemed real by the discriminator. To apply GAN to video prediction, \cite{CGAN} imposes conditional losses to ensure the images are not only realistic but also similar to the previous frames. 

Video interpolation is closely related to video prediction, and can help increase frame rate and generate slow motion. Unlike prediction, video interpolation can take advantage the future frames in addition to the previous frames. Refer to \cite{video-int-survey} for a comprehensive survey. It classifies the existing approaches into flow-based, CNN-based, phase-based, GAN-based, and hybrid methods. 


Video prediction and interpolation both require estimating each pixel value, which is inherently very challenging. For the purpose of depth estimation, our goal is to estimate the collision time. Therefore, we interpolate the target's trajectory instead of every pixel in a video frame.

\para{Audio visual learning:} Human perceives the world through multiple modalities, such as vision, hearing, touch, smell, touch, and taste. To empower machines with human-like capabilities, a recent trend is to leverage multimodality learning. Among them, visual and audio are the most important modalities and has attracted significant amount of work on visual-audio learning. As shown in \cite{audio-visual-survey}, the existing work in this area can be classified into audio-visual correspondence learning, audio-visual separation (\eg, using visual scenes to facilitate audio separation), audio-visual localization (\eg, identifying the sound source in the visual scene), audio-visual generation, and audio-visual representation learning.

Our work is built on existing work on audio-visual correspondence learning, in particular, cross-modality retrieval, which uses samples from one modality to retrieve samples in another modality. \cite{obj-sound} maps two modalities into a joint embedding space and determines correspondence based on their cosine similarity. \cite{audio-visual} develops a cross-modality self-supervised method and curriculum learning to improve the efficiency and performance. We use SoundNet and ResNet to extract audio and visual features from each modality and develop our own DNN to identify the pixels in the images corresponding to the collision sound. 

\para{Summary:} Our work is inspired by a number of existing works and incorporates some of the techniques proposed earlier, including audio and video feature extraction. Different from the existing work, \ours is the {\em first} that leverages "flash-to-bang" phenomenon and the property of collisions to detect events in the audio and video streams for depth estimation.

%% file: limit.tex
\section{Limitations and Potential Solutions}

Our work is the first that applies flash-to-bang notion to enable passive depth estimation. There are a few areas to further improve its performance and applicability. 

First, our current approach leverages the impact sound for depth estimation. While many scenarios involve impact sound as mentioned earlier, there are scenarios that do not contain sharp impact sounds. For example, the target is static or silent,  the target is a user who is speaking, or the target has a continuous movement (\eg, a running engine). There are a few ways to generalize our approach to more contexts. First, if the video frame contains an object that generates impact sound, we can apply our approach to estimate its depth and then use its estimated depth to infer the depth of its nearby objects (\eg, by combining with the monocular camera based depth estimation). Second, for targets that generate sound but not sharp impact (\eg, speech or continuous motion), it may be feasible to learn the association between the sound and movement (\eg, mouth movement vs. sound that is produced) and apply flash-to-bang concept.

Second, FBDepth has higher accuracy over a large distance. To improve its accuracy at a short distance, we can also combine with monocular camera based solutionsc, which yields good accuracy at a short distance. The existing monocular solution works well when the testing data are similar to the training data. FBDepth can complement monocular solution by generalizing a reasonable estimate for arbitrary scenes and using the monocular method to refine.

%% file: conclusion.tex
\section{Conclusion}
\label{sec:conclusion}

In this paper, we develop a novel depth estimation method based on the "flash-to-bang" phenomenon. By aligning the video with the audio and detecting the events from both, we can estimate the depth in the wild without calibration or prior knowledge about the environment or target. Our extensive evaluation shows that our approach yields similar errors across varying distances. In comparison, the errors of several existing methods increase rapidly with distance. Therefore, our method is particularly attractive for large distances (\eg, beyond 5 m). As part of our future work, we are interested in further enhancing the accuracy of our method, generalizing to more contexts, and using the estimated depth to the collision to estimate the depth to other objects in the scene.